\documentclass[runningheads]{llncs}
\usepackage{graphicx}

\begin{document}
\title{Big Data Challenges of FAST}
\author{Youling Yue\inst{1,2}, Di Li\inst{1,2,3}}
\authorrunning{Yue et al.}
\institute{
National Astronomical Observatories, Chinese Academy of Sciences,  Beijing 100101, China, P.R.
\and
CAS Key Laboratory of FAST, NAOC, Beijing, 100101, China, P.R.
\and
University of Chinese Academy of Sciences, Beijing 100049, China, P.R.
\email{dili@nao.cas.cn} }
\maketitle              % typeset the header of the contribution
\begin{abstract}
We present the big-data challenges posed by the science operation of the Five-hundred-meter Aperture Spherical radio Telescope (FAST). Unlike the common usage of the word `big-data', which tend to emphasize both quantity and diversity, the main characteristics of FAST data stream is its single-source data rate at more than 6 GB/s and the resulting data volume at about 20 PB per year. We describe here the main culprit of such a high data rate and large volume, namely pulsar search, and our solution.

\end{abstract}
\section{Introduction}
The Five-hundred-meter Aperture Spherical radio Telescope (FAST)
(Nan et al. 2011)
%\cite{Nan2011ijmpd}
is a Chinese mega-science project, originated from a concept for the Square Kilometer Array (SKA). Once the SKA concept moves toward small-dish-large-number, FAST becomes a stand-alone project, which was formally funded in 2007. The construction begun in 2011 and finished in 2016. The ensuing commissioning phase will come to an end in September, 2019.

With an unprecedented gain of $2000~\mathrm{m^2/K}$, superseding any antenna or antenna array ever built, FAST poises to increase volume of knowledge, i.e. the known sources and pixels of sky images, of pulsar and HI, by one order of magnitude. To accomplish these goals in an efficient manner, a Commensal Radio Astronomy FAST Survey (CRAFTS; Li et al. 2018) has been designed and tested. Utilizing the FAST L-band Array of 19-beams (FLAN) in drift-scan mode, CRAFTS aims to obtain data streams of pulsar search, HI imaging, HI galaxies, and transients (Fast Radio Burst--FRB, in particular), simultaneously. Our own innovative technologies have made such commensal observation feasible, which was not the case in past at, e.g. GBT, Arecibo, Parkes, etc. CRAFTS requires acquiring and processing data in four conceptually independent backends. The fast-varying nature of pulsar signals demand a sampling rate of 10000 Hz or more, which makes pulsar search the main component of FAST's big-data volume.
FAST is thus facing a big data challenge comparable to that of Large Synoptic Survey Telescope (LSST), which also expects data over 100 PB.

We quantify the data requirement of CRAFTS, pulsar search in particular, in section 2, describe our current data solution in section 3, and discuss our future plan and challenges in the final section.

\section{CRAFTS pulsar data challenge}
The golden standard of pulsar surveys is the Parkes multi-beam pulsar survey (PMPS) (Manchester et al. 2001) started in  August 1997. PMPS discovered over 800 Pulsars.
It utilizes 250 $\mu$s 1 bit sampling, 96 frequency channels, 288 MHz bandwidth, 13 beams, amounting to a 0.64 MB/s data rate.
The total data volume is about 4 TB, which is trivial for today's desktop.
%Along with other surveys, Parkes telescope discovered around half of the $\sim$ 2600 yet known Pulsars.

In the two decades following the PMPS, the computing power available has grown exponentially. So does the appetite of astronomers, even outpace the industrial trend, just like  Otto in ``A Fish out of Walter''.
Assuming an increase in pulsar search data volume by 100 fold in 10 years, we should expect 6.4GB/s data rate and 40 PB total volume for FAST pulsar survey. Forecasting technology requirements on decades time scale is necessarily unreliable, especially when Moore's law seems to be hitting the quantum limit in recent years.  These numbers, however, are amazingly close to the reality of FAST.
%detailed estimation below.

FLAN has 19 beams, with each has 2 polarization of 500 MHz bandwidth, generating a total of 38 analog signals.
With 1 GHz 8-bit sampling, this leads to 38 GB/s total raw data rate.
%400MHz bandwidth 50 MHz sideband to leave band for cutting  aliasing.
%1GHz sampling rate 8bit.
Store raw data is not an economically feasible solution,
not to mention the processing pressure.
The solution is shrink the data volume with acceptable loss.

\section{CRAFTS data processing}

There are two main challenges of CRAFTS pulsar survey. One is to store, transfer, and process the data. The other is to sift through billions of candidate signals to find about 1000 new pulsars. AI techniques are now common in such tasks.

Pulsar data can be simplified as a 2D matrix or a image, usually 8-bit.
Pulsar signal is repeating with characteristic patterns, but also with much variation. An example is given in fig. 1.

\begin{figure}
\includegraphics[width=\textwidth]{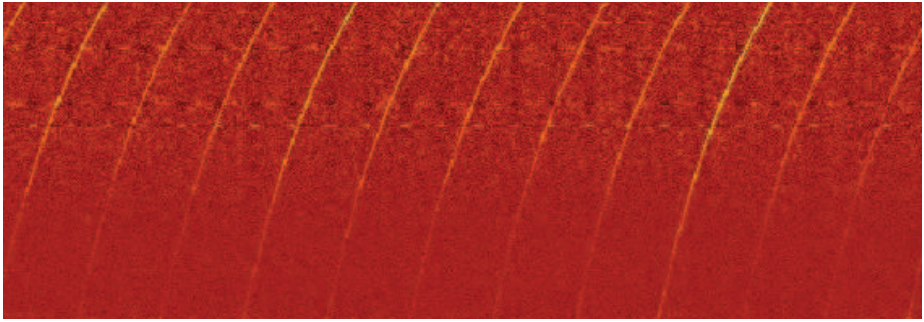}
\caption{A sample image of pulsar data. Each bright line a pulsar signal. This is a strong pulsar. Most pulsar signal is weaker than noise, thus need to be folded according to the period to be shown.} \label{fig1}
\end{figure}

The processing of pulsar data is a multi-dimensional problem.
Each dimension is computation of O(n) or O(n*log(n)).
Since most pulsar signal is week, we need to use FFT to find the period.
For pulsar in binaries, the pulsar period is not exact the same, it is modulated by the orbital period. This add 4 dimensions of freedom.
Searching will be computing intensive.
Each dimension  add the computing complexity by a factor of 100 to 1000.
This increase the computing of PFLOPS at least by a factor of $10^8$, i.e. $10^{23}$ FLOPS (0.1 YFLOPS).

First step of solving storage and transfer problem is data compression.
With FLAN, the on-board processing provided by ROACH2, facilitates 4k channels, full polarization, 8 bit sampling, and time resolution of 8 $\mu$s. The fastest pulsar has period around 1 ms.
Time resolution of 8 $\mu$s is overkill, 0.05-0.1 ms is enough and leaves some room for submillisecond pulsars.
By compressing to time resolution $\sim$50 $\mu$s,
we cut the data rate to 1/6 of the raw, thus 6.3 GB/s or 544 TB/day(24 h), which can be transferred through 100GbE.
This means around 200 PB/yr first stage data,
which is still huge. It must be further compressed after the first round processing.
Being irreversible, the second stage of compression has to be done with caution and necessary compromise.
The current consideration is to reduce the channel number from 4k to 1k and data sampling 8 bit to 2 bit, a further compression factor of 16 can be achieved, corresponding to $\sim$12 PB per year.
Considering maintenance time and other science data, FAST will store about 7-10 PB pulsar data per year.

%But there is case like FRB, the signal is not repeating

\section{Discussion and conclusion}
For single-dish telescopes, especially FAST, their data volumes are manageable by modern technology standards.  The bigger challenge lies in optimizing the  solution under  budget constraints.

In data processing, pulsar search has several unsolved problems. Searching for pulsar in binaries, e.g., requires over ZFLOPS computational power, close to YFLOPS.
If exponential growth, even one slower than Moore's law, still holds,
pulsar search problem for FAST can be solved after 20 years or more.
By then though, much larger pulsar survey will take place, e.g.\ phased array receiver with 100 beams or more for FAST, SKA surveys, and etc. Emerging techniques such as quantum computing, AI, new algorithms will also see wider application.

\section{Acknowledgement}
This work is supported by National Key R\&D Program of China grant No. 2017YFA0402600, the National Natural Science Foundation of China grant No. 11725313 and  CAS ``Light of West China" Program.


\end{document}